\documentclass[11pt]{article}
\usepackage[utf8]{inputenc}
\usepackage{graphicx}
\usepackage{amsmath, amssymb}
\usepackage{booktabs}
\usepackage{geometry}
\usepackage{hyperref}
\title{Multimodal Non-Semantic Feature Fusion for Predicting Segment Access Frequency in Lecture Archives}

\author{
Ruozhu Sheng\textsuperscript{1,2,*} \and
Jinghong Li\textsuperscript{1,2} \and
Shinobu Hasegawa\textsuperscript{3,*} \\
\\
\textsuperscript{1} Graduate School of Advanced Science and Technology, \\
Japan Advanced Institute of Science and Technology, Ishikawa, Japan \\
\textsuperscript{2} J-MAX Co., Ltd., Kamiishizu, Ogaki, Gifu, Japan \\
\textsuperscript{3} Center for Innovative Distance Education and Research, \\
Japan Advanced Institute of Science and Technology, Ishikawa, Japan \\
\texttt{sheng@jaist.ac.jp, hasegawa@jaist.ac.jp}
}

\date{} 

\begin{document}

\maketitle

\begin{abstract}
This study proposes a multimodal neural network-based approach to predict segment access frequency in lecture archives. These archives, widely used as supplementary resources in modern education, often consist of long, unedited recordings that make it difficult to keep students engaged. Captured directly from face-to-face lectures without post-processing, they lack visual appeal. Meanwhile, the increasing volume of recorded material renders manual editing and annotation impractical. Automatically detecting high-engagement segments is thus crucial for improving accessibility and maintaining learning effectiveness. Our research focuses on real classroom lecture archives, characterized by unedited footage, no additional hardware (e.g., eye-tracking), and limited student numbers. We approximate student engagement using segment access frequency as a proxy. Our model integrates multimodal features from teachers’ actions (via OpenPose and optical flow), audio spectrograms, and slide page progression. These features are deliberately chosen for their non-semantic nature, making the approach applicable regardless of lecture language. Experiments show that our best model achieves a Pearson correlation of 0.5143 in 7-fold cross-validation and 69.32\% average accuracy in a downstream three-class classification task. The results, obtained with high computational efficiency and a small dataset, demonstrate the practical feasibility of our system in real-world educational contexts.
\end{abstract}


\section{Introduction}

Online learning originated in the 19th century through correspondence education and has evolved significantly in today's digital age through advances in computer and Internet technologies. The emergence of Open Education Resources (OER) and Massive Open Online Courses (MOOCs) has fundamentally transformed educational accessibility \cite{saykili2018distance}. While the trend toward online education was already growing, the COVID-19 pandemic accelerated this transformation to unprecedented levels, with UNESCO reporting approximately 862 million students, almost half of the world's student population, affected by school closures across 107 countries \cite{abuhammad2020barriers}. This global shift prompted higher education institutions worldwide to rapidly adopt online learning platforms. Even after the pandemic has subsided, most institutions continue to rely on widely adopted online learning platforms, making it increasingly important to enhance the accessibility and effectiveness of recorded lecture archives.

Among the various formats of recorded educational videos, lecture archives, which are complete recordings of face-to-face lectures without editing, have emerged as a prevalent choice for many institutions. This popularity stems primarily from their cost-effectiveness and ease of distribution. The Japan Advanced Institute of Science and Technology (JAIST) exemplifies this approach, having systematically recorded face-to-face lectures through their Learning Management System since 2006, thereby creating an extensive archive for supplemental learning \cite{hasegawa2007}. However, these unedited, long-form recordings present two significant challenges. First, students often find it difficult to maintain attention throughout extended viewing periods \cite{guo2014video}. Second, instructors face substantial difficulties in manually editing or managing large volumes of video content. 

Identifying high engagement segments is considered a promising way to improve the accessibility and effectiveness of lecture archives. However, it is difficult to directly measure student engagement with instructional videos without relying on methods such as eye tracking or user self-reporting. These approaches are often impractical in real-world educational environments~\cite{guo2014video}. To address this challenge, prior studies have introduced behavioral indicators derived from interaction data. For example, Guo et al. proposed the use of engagement time—the duration students spend watching a video—as a proxy for interest.

Building upon this idea, Bulathwela et al.~\cite{bulathwela2020vlengagement} introduced the VLEngagement dataset, which estimates video-level engagement by computing a normalized viewing duration aggregated across large numbers of users. Specifically, they defined Engagement Score = Average Watch Time / Video Duration, providing a cost-effective and scalable labeling method for large collections of educational videos. However, the engagement labels in this dataset are defined at the whole-video level, resulting in a coarse granularity that limits its applicability to tasks such as segment-level attention modeling or highlight extraction.

Inspired by these approaches, we propose to use segment access frequency as a more fine-grained and context-appropriate measure of engagement. This metric, calculated from the number of playback events associated with each time segment, provides a practical and scalable solution that does not depend on semantic content or specialized hardware. It is particularly suitable for classroom lecture recordings, which typically lack rich annotations or auxiliary sensors.

This study focuses on lecture archives recorded in real classroom environments. These recordings are typically unedited, collected without auxiliary hardware such as eye-tracking devices, and involve only a small number of learners. In many cases, the recording setup includes a microphone and camera mounted on the ceiling at some distance from the instructor, resulting in limited audio quality. Although students can hear and understand the content during playback, the recordings are often too noisy or indistinct for reliable automatic transcription. One representative example of such a setting is the video archive system accumulated at JAIST, where face-to-face lectures are routinely recorded and made available through the institutional LMS. We aim to develop a lightweight and efficient prediction method based on non-semantic multimodal features to address the challenges of such resource-limited settings. This approach is designed to avoid semantic dependence, enable cross lingual adaptability, and minimize training costs. Furthermore, by generating engagement labels automatically from aggregated playback data, the system is intended to support scalable deployment in practical educational contexts.

Building on these considerations, this study is structured around the following research questions and the associated contributions of our work:

\begin{quote}
    \textbf{RQ1:} Can non-semantic visual and audio features extracted from lecture archives predict segment access frequency with meaningful accuracy, even under limited data conditions?
\end{quote}

\begin{quote}

    \textbf{RQ2:} RQ2: For the non-semantic features identified in RQ1, how do different multimodal fusion strategies affect the performance and efficiency of prediction models?

\end{quote}

\begin{quote}
  \textbf{RQ3:} What is the relative contribution of each modality (action, voice, slide) to the overall prediction performance, and which backbone architecture provides the best trade-off between accuracy and computational cost?
\end{quote}

\vspace{1em}
To address these research questions, this study makes the following contributions:

\begin{itemize} 
    \item We introduce a lightweight and language-independent prediction framework that estimates segment access frequency based solely on observable multimodal signals, without relying on semantic understanding or specialized equipment. 
     
    \item We design and compare two alternative strategies for integrating multimodal information and confirm the advantage of that early feature fusion achieves higher prediction accuracy and training efficiency in resource-limited scenarios. 
     
    \item We conduct a comprehensive ablation and backbone analysis to evaluate the relative contribution of different modalities and identify the optimal model structure for this task. Our results highlight the central role of action features and confirm the effectiveness of ResNet-based architectures under real-world constraints. 
\end{itemize} 


\section{Literature review}
Our research is situated at the intersection of two areas: video summarization and student engagement modeling. While video summarization focuses on selecting key content segments, engagement modeling aims to estimate which parts of a lecture attract student attention. This section reviews representative works from both directions to clarify the foundation and scope of our non-semantic, engagement-driven approach.

\subsection{Video Summarization}

Modern educational platforms offer learners extensive access to recorded lecture videos, prompting the need for technologies that can enhance the efficiency of video-based learning~\cite{benedetto2024abstractive}. Video summarization addresses this need by enabling educators and students to quickly discern the educational value within lengthy recordings~\cite{sablic2021video}. It generates concise representations of video content through combinations of still images, short segments, visual diagrams, or textual annotations. Early research proposed rule-based methods for extractive summaries, typically selecting representative keyframes, segments, or transcript snippets~\cite{dimitrova2004context, alaa2024video}.

With advancements in machine learning, deep learning models have gained prominence in video summarization, particularly for capturing temporal dependencies in time-series data~\cite{han2019review}. Recent studies have developed sophisticated approaches tailored to diverse video types. For instance, Singh et al.~\cite{singh2024bayesian} introduced a deep learning framework integrating Bayesian fuzzy clustering with a Deep Convolutional Neural Network (Deep CNN), optimized via a hybrid Lion-Deer Hunting (LDH) algorithm. Their approach achieved significant improvements in precision, recall, and F1-score on crowd video datasets. However, it is designed for dynamic scenes with dense motion and clear separation of foreground and background, which contrasts sharply with the static camera angles, minimal motion, and subtle engagement cues typical of lecture archives. These differences necessitate distinct design considerations for educational contexts.

On the other hand, Kim et al. emphasize the need to move beyond traditional semantic or content-only analyses, advocating for the use of interaction data to optimize video design and highlighting the pedagogical value of editing videos for brevity and improved engagement~\cite{10.1145/2556325.2566237}.

Various time-series models leveraging recurrent architectures have been explored for video summarization. Agyeman et al.~\cite{agyeman2019soccer} developed a hybrid model combining three-dimensional Convolutional Neural Networks (3D-CNN) with Long Short-Term Memory (LSTM) layers to classify events in soccer videos, achieving 96.8\% accuracy. While effective for sports and surveillance videos with rapid scene changes, such models are less suitable for lecture archives, which feature limited visual variation, sparse motion, and extended durations. Moreover, the high computational cost of training recurrent neural networks (RNNs) on long videos poses practical challenges for resource-constrained educational settings. To improve summarization performance without heavy temporal modeling, some recent methods have focused on enhancing input representation: Tan et al.~\cite{10.1145/3661725.3661781} used adaptive clustering to extract more meaningful keyframes, while Khan et al.~\cite{khan2024deep} introduced multi-scale feature maps to capture both detailed and semantic-level information.

In the education field, Andra and Usagawa~\cite{andra2019automatic} summarized lecture videos through an Attention-based Recurrent Neural Network (RNN) that combines segmentation with the summarization process. The RNN architecture generates a natural summary by capturing critical words and conveying a lecture's central message through attention-based weighting and linguistic features. However, their method significantly depends on semantic analysis of lecture content. Such transcripts are not always available for lecture archives, and the accuracy of existing Automatic Speech Recognition (ASR) techniques is not always satisfactory for audio data with noise and precise terminology. Moreover, the limited availability of annotated data for lecture videos poses another challenge for such methods.

To address the issue of limited annotated data, Vimalaksha et al.~\cite{vimalaksha2018automated} provided a mechanism to segment lecture videos into multiple parts based on crowdsourcing. While this approach alleviates the annotation problem to some extent, manual crowdsourcing methods have their own limitations. They require real-time recording, are labor-intensive, and have strict time synchronization requirements, making them prone to bias. Therefore, despite the contribution of crowdsourcing methods, there remains a need for more generalized approaches to effectively tackle the annotation problem for lecture videos, particularly in contexts with limited resources and multilingual content as addressed in our current research.

\subsection{Student Engagement and Attention}
Student engagement is crucial in higher education, yet fostering active participation in online learning environments remains a persistent challenge ~\cite{vermeulen2024promoting}. To address this issue, researchers have examined various instructional and content-related factors that may influence how students interact with lecture materials.

For instance, embedded semantic approaches have been developed to enhance student interaction within classroom settings. Deng et al. explores how embedding questions within pre-class instructional videos influences learners’ experiences and outcomes in a flipped classroom context ~\cite{deng2024effects}. However, semantic analysis primarily focuses on textual content and overlooks learner interactions with the video, such as pausing, rewinding, or skipping sections. These behaviors provide crucial insights into attention patterns and engagement levels, which semantic methods cannot capture.

On the other hand, several studies have explored the design of video lectures without treating semantic features as a key element for promoting engagement. For example, Chen et al. investigated the impact of different video lecture formats on learning outcomes, cognitive load, and emotional responses ~\cite{chen2015effects}. Their findings suggest that featuring instructors on screen not only enhances students’ sense of connection but also reduces cognitive overload. This highlights the importance of visual presence and presentation style in maintaining student attention and fostering participation in online learning environments.

With regard to visual presence and presentation style, Shi et al. examined how instructors’ visual attention and lecture delivery styles influence students’ perceived engagement and academic performance across various instructional formats ~\cite{shi2024impacts}. Similarly, Zhang et al. employed eye-tracking and visualization technologies to investigate the effects of different instructional delivery styles on student viewing behavior~\cite{zhang2018effects}. Their analysis showed that students were more responsive to auditory cues—such as pauses and vocal emphasis—than to visual elements like gestures or slide transitions. Collectively, these studies underscore the importance of instructor-related features—both visual and auditory—in shaping student engagement with lecture content. In particular, they highlight the pivotal role of auditory cues in sustaining attention during recorded lectures, even when visual presence is emphasized. These findings suggest that integrating non-semantic instructional features, such as visual presence and vocal modulation, is instrumental in directing learners’ attention toward key concepts in a lecture.

In addition to instructor behavior, students' behavioral data has also been used to examine engagement at scale. Kim et al. ~\cite{kim2014understanding} conducted a large-scale analysis of video interaction patterns in MOOCs and identified several recurring behaviors—such as replaying specific moments and revisiting explanation-heavy segments—as signals of focused engagement. Such naturally occurring patterns provide a foundation for scalable engagement modeling based on interaction data, without requiring semantic understanding or manual annotation.

Inspired by these findings, we incorporate instructor-related features, such as action and voice, together with playback behavior statistics, to support the prediction of high-engagement segments in real-world lecture archives.

\subsection{The position of this work}

Our research focuses on estimating student engagement levels in lecture videos by analyzing non semantic multimodal features. Unlike most existing approaches that rely on semantic content analysis, we emphasize how students interact with lecture materials through their actual viewing behaviors. This approach offers three main advantages: it is applicable across different languages as it does not require semantic understanding, it reduces the need for manual annotation by utilizing access logs, and it identifies the lecture segments that exhibit higher levels of student engagement. By combining features extracted from instructor actions, voice, and slides with patterns of student interaction, our method provides a lightweight and flexible solution for engagement analysis in real educational environments. 


\section{Research Design and Methodology}

\begin{figure}[htbp]
\centering
\fbox{
\includegraphics[width=9cm]{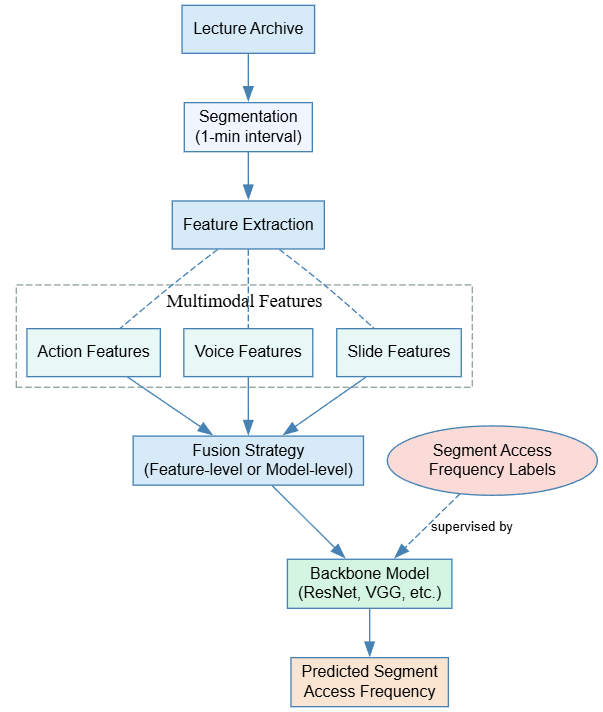}
}
\caption{System pipeline for predicting segment engagement from lecture archives.}
\label{fig:pipeline}
\end{figure}

\vspace{1mm}
This study proposes a lightweight and scalable system to predict student engagement patterns in real classroom lecture archives. The overall workflow, illustrated in Figure~\ref{fig:pipeline}, consists of four main stages. First, the lecture archives are segmented into uniform time intervals. Second, multimodal features that do not rely on semantic content—including instructor actions, voice spectrograms, and slide transitions—are extracted from each segment. Third, these features are fused and processed by a deep neural network model. Finally, the model outputs predictions of segment access frequency, supervised by labels automatically generated from aggregated access logs. Notably, the entire pipeline from feature extraction to label generation can be fully automated without manual intervention. The system is designed to operate efficiently in resource-constrained environments without relying on semantic analysis or specialized hardware.

\subsection{Dataset}

\subsubsection{Lecture archives}
\begin{figure}[htbp]
\includegraphics[width=14 cm]{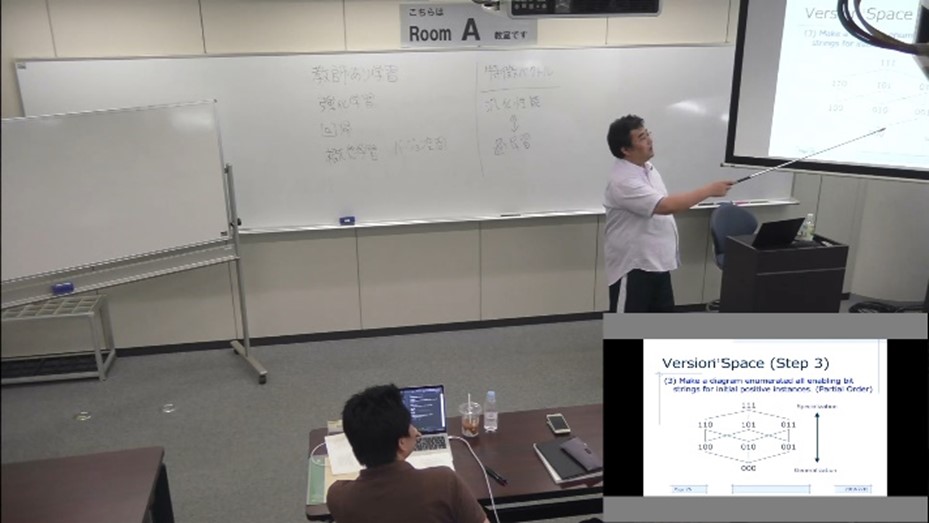}
\centering
\caption{Original Lecture Archive, I239 Machine Learning.}
\label{fig1}
\end{figure}

The lecture archives used in this research were recorded from the I239 Machine Learning course offered through the JAIST Learning Management System (JAIST-LMS). The course consisted of seven lessons, each recorded to support students' reflection and supplemental learning following face-to-face instruction. A ceiling-mounted camera with a fixed angle and a ceiling microphone captured both the lecturer’s and students’ voices. The video files were recorded at 1920×1080 resolution and 30 frames per second. Each lesson lasted approximately 100 minutes. The archives included the podium area, whiteboard, and instructor. In addition, the slide content was integrated into the right-bottom corner of the archives, as shown in Fig.~\ref{fig1}.\\

\vspace{1mm}
\subsubsection{Label Generation}

The JAIST-LMS extends Video.js to track students’ access to specific durations of lecture archives. Based on this detailed playback history, we generated student engagement labels using segment access frequency, defined as the number of times each one-minute segment was accessed across all users.

To reduce noise and ensure data reliability, we first excluded all raw viewing records shorter than one minute. We then computed the total valid viewing time per student and retained only those students who watched more than five minutes of a given lecture, labeling them as valid viewers. Segment access frequencies were calculated by aggregating only these valid viewers’ logs. Table~\ref{tab:viewing_stats} summarizes the number of valid viewers and the total valid viewing time for each lecture.

Table~\ref{tab:viewing_stats} shows the statistics of valid viewers and total valid viewing time per lecture.\\

\begin{table}[htbp]
\centering
\caption{Valid viewers and total viewing time for each lecture.}
\label{tab:viewing_stats}
\small
\begin{tabular}{lcc}
\toprule
\textbf{Lesson} & \textbf{Valid Viewers} & \textbf{Total Viewing Time (min)} \\
\midrule
Lesson-1 & 10 & 532 \\
Lesson-2 & 6 & 605 \\
Lesson-3 & 11 & 647 \\
Lesson-4 & 8 & 408 \\
Lesson-5 & 8 & 607 \\
Lesson-6 & 9 & 643 \\
Lesson-7 & 9 & 939 \\
\midrule
\textbf{Average} & 8.71 & 625.86 \\
\bottomrule
\end{tabular}
\end{table}

\vspace{1mm}
After data filtering, Table~\ref{tab:viewing_stats} shows that our label dataset is extremely limited in size. On average, each lecture has only 8.71 valid viewers. This is likely because students had already attended the face-to-face class sessions, and the archives were mainly used as a review resource. In such cases, students may choose to revisit only selected parts of the lecture or rely on other materials, such as textbooks or slides, for review. As a result, the overall number of archive viewers is low.

However, this type of review behavior may also indicate stronger learning intent. According to Kim et al. \cite{kim2014understanding}, students who engage in repeated viewing tend to show more focused and high-peak interaction patterns compared to first-time viewers. These behaviors are often goal-driven, as students selectively locate and revisit important parts of the content. Therefore, although our dataset is small, the engagement signals it captures may be more concentrated and meaningful, offering a reliable proxy for identifying high-engagement segments.

To further adapt the data for model training, we removed the non-instructional portions at the beginning and end of each video. All lecture archives were trimmed or padded to a standardized length of 95 minutes, focusing exclusively on instructional content. Each archive was then divided into one-minute segments. For each segment, we calculated the segment access frequency by summing the number of valid viewers who accessed it.

To suppress short term fluctuations, a centered moving average with a five segment window was applied to the raw frequency sequence. The resulting values were then normalized within each lecture by dividing by the maximum segment frequency, yielding engagement labels in the range of [0, 1]. In total, we obtained 665 labeled segments across the seven lectures for subsequent model training and evaluation. The segment access frequency labels were automatically generated from raw csv format access logs through a batch-processing script, eliminating the need for manual annotation.

\subsubsection{Evaluation Metrics}
\label{sec:evaluation_metrics}

To assess the performance of our proposed system, we adopt both regression-based metrics and 3-classification accuracy for a comprehensive evaluation.

\textbf{Regression metrics.} Since the primary task is to predict a continuous attention level, represented by normalized segment access frequency values between 0 and 1, we employ the following standard metrics:\\
\begin{itemize}
    \item \textbf{Mean Squared Error (MSE):} Measures the average squared difference between the predicted and ground-truth values.
    \item \textbf{Mean Absolute Error (MAE):} Computes the average magnitude of absolute prediction errors.
    \item \textbf{Coefficient of Determination ($R^2$):} Indicates the proportion of variance in the ground truth that is explained by the predictions.
    \item \textbf{Pearson Correlation Coefficient (PCC):} Evaluates the linear correlation between predicted and ground-truth sequences, reflecting trend similarity.
\end{itemize}
\vspace{1mm}

\textbf{3-classification accuracy.} For interpretability and visualization, we further discretize the continuous predictions into three attention zones:\\

\begin{itemize}
    \item \textbf{High attention:} Segment access frequency $> 0.5$
    \item \textbf{Medium attention:} Values in between
    \item \textbf{Low attention:} Segment access frequency $< 0.2$
\end{itemize}

\vspace{1mm}
We then compute the accuracy of the 3-classification as the proportion of segments where the predicted and ground-truth zones match. This discrete representation supports downstream visualization, such as attention heatmaps and facilitates understanding of the predicted attention distribution in lecture archives.

\subsection{Feature Extraction and Preprocessing}

\subsubsection{Action Features (A)}
According to the previous study by Zhang et al. \cite{zhang2018effects}, the behavior of the instructor influences the attention of students. Therefore, we obtained the instructor's action in the archive segments by the optical flow \cite{burton1978thinking}, the pattern of apparent motion of objects, surfaces, and edges in each segment caused by the relative motion between observer and scene.

However, the optical flow could not work well because the corner point is always generated on the slide rather than the instructor in a default setting. To solve this problem, we first mask the students' seating area and then capture the instructor's body structure feature by OpenPose \cite{cao2019openpose}, the first open-source real-time system available to detect multi-person 2D poses, including body, feet, hands, and facial key points. Next, we calculate the optical flow for the instructor's action based on the captured body structure. This research uses the Lucas-Kanade method to calculate the optical flow for every segment \cite{lucas1981iterative}. Fig.~\ref{fig:action_features} shows an action feature map from a one-minute archive segment extracted by our method.

\subsubsection{Voice Features (V)}
\begin{figure}[htbp]
\centering
\begin{minipage}{0.48\linewidth}
    \centering
    \includegraphics[width=\linewidth]{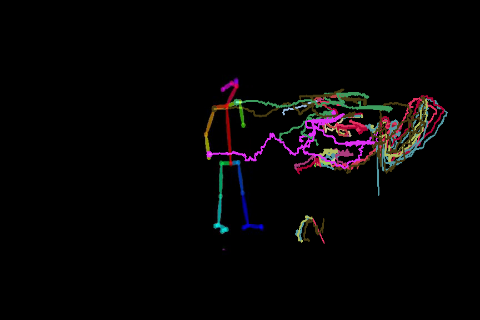}
    \caption{Action Feature.}
    \label{fig:action_features}
\end{minipage}
\hfill
\begin{minipage}{0.48\linewidth}
    \centering
    \includegraphics[width=\linewidth]{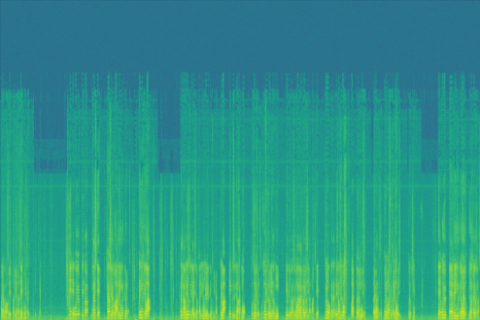}
    \caption{Voice Feature.}
    \label{fig:voice_features}
\end{minipage}
\end{figure}

According to a previous study by Wyse \cite{wyse2017audio}, neural networks used in classification or regression can benefit from spectrograms which are a visual representation of the spectrum of signal frequencies as it varies with time \cite{flanagan1972speech}. In addition, they retain more information than most hand-crafted features traditionally used to analyze voice or sound and have a lower dimension than raw data.
We implemented a SciPy-based approach utilizing the spectrum function to generate spectrograms from the lecture audio. The spectrograms were computed using Fast Fourier Transform (FFT) with a window size of 1024 samples and an overlap of 128 samples between adjacent windows at a 44.1kHz sampling rate. This configuration provides sufficient frequency resolution while maintaining temporal precision necessary for analyzing speech patterns in lecture recordings. The spectrogram generation process effectively converts the time-domain audio signal into a two-dimensional time-frequency representation, capturing both temporal and frequency characteristics of the instructor's voice. The resulting spectrograms were then processed as feature maps for our deep learning model, as shown in Fig.~\ref{fig:voice_features}. This approach enables our model to learn from both the frequency content and temporal dynamics of the instructor's speech, while maintaining computational efficiency.

\vspace{1mm}
\subsubsection{Slide Features (S)}

Slide transitions indicate lecture pacing and content structure, potentially influencing student engagement. To capture this, we extract numerical features representing net slide progression, processed through a five-step pipeline to ensure temporal alignment and compatibility with other modalities.

The processing pipeline consists of five steps:

\begin{enumerate}
    \item \textbf{Net progression:} For each 5-minute segment $i$, we compute the net forward movement as:
    \[
    P_{\text{raw}}[i] = \max\{x_i\} - \max\{x_{i-1}\}, \quad \text{where } P_{\text{raw}}[i] = 0 \text{ if the difference is negative or zero.}
    \]
    \item \textbf{Temporal resolution adjustment:} Each $P_{\text{raw}}[i]$ is repeated 5 times to form a 1-minute resolution sequence:
    \[
    P_{\text{1min}} = \text{repeat}(P_{\text{raw}}, 5)
    \]
    \item \textbf{Smoothing:} Apply a moving average filter with window size 5:
    \[
    P_{\text{smooth}}[i] = \frac{1}{5} \sum_{j=i-2}^{i+2} P_{\text{1min}}[j]
    \]
    \item \textbf{Normalization:} Within each lesson, normalize the values by the lesson-wise maximum:
    \[
    P_{\text{norm}}[i] = \frac{P_{\text{smooth}}[i]}{\max(P_{\text{smooth}})}
    \]
    \item \textbf{Matrix construction:} Each normalized value is scaled to 8-bit and expanded into a uniform 2D matrix:
    \[
    S[i] = \mathbf{1}_{h \times w} \cdot \text{int}(P_{\text{norm}}[i] \times 255)
    \]
    where $\mathbf{1}_{h \times w}$ denotes a matrix of ones with spatial dimensions matching other modalities.
\end{enumerate}

This design ensures that the slide feature is temporally aligned and dimensionally compatible with the action and voice feature matrices for multimodal fusion.

\subsubsection{Temporal Smoothing}

To enhance the temporal consistency of predictions and suppress short-term fluctuations, we apply smoothing techniques to the attention level sequences produced by the regression model. These smoothed values are subsequently used for interpretation, visualization, and threshold-based classification.

Let $y[i]$ denote the predicted attention level at minute $i$. We apply the following three smoothing methods:

\begin{itemize}
    \item \textbf{Moving Average:} A centered rolling mean applied over a window of 5 minutes:
    \[
    \hat{y}[i] = \frac{1}{w} \sum_{j = i - \lfloor w/2 \rfloor}^{i + \lfloor w/2 \rfloor} y[j], \quad w = 5
    \]
    This method is simple yet effective in suppressing short-term fluctuations.

    \item \textbf{Savitzky–Golay Filter:} Implemented with a window length of 7 and a second-order polynomial, this filter performs local polynomial regression to preserve peak shapes while smoothing the signal.

    \item \textbf{Kalman Filter:} A one-dimensional recursive Bayesian estimator using the FilterPy library. We set the state transition and observation matrices as $F=H=[1]$, initial covariance $P=500$, measurement noise $R=0.05$, and process noise $Q=10^{-4}$.
\end{itemize}

\vspace{1mm}
These methods are applied post hoc to the predicted sequences and do not affect the model training process. Among them, the moving average achieves the best trade-off between simplicity and performance in our experiments. Detailed comparisons are presented in Section~\ref{tab:action_smoothing}. The smoothed sequences are also used to derive three-class attention zones via thresholding, as described in Section~\ref{sec:evaluation_metrics}.

\subsection{Model Architecture and Experimental Settings}

\subsubsection{Feature Fusion Strategies}

\begin{figure}[htbp]
\centering
\includegraphics[width=7cm]{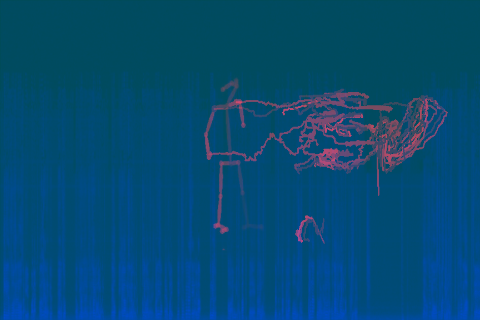}
\caption{Feature fusion input example combining action, voice, and slide features into RGB channels.}
\label{Fsample_AVS}
\end{figure}

\begin{figure}[htbp]
\centering
\fbox{
\includegraphics[width=14 cm]{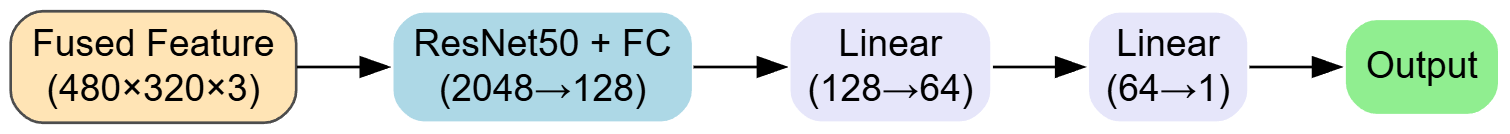}
}
\caption{Process of feature-level fusion (example based on ResNet).}
\label{feature_stacking}
\end{figure}

We prepare two deep learning architectures to detect the focal points of the above mentioned features.
The first option, called "Feature Stacking," converts different feature maps of each archive segment into RGB channels of a single image file. An example of such a stacked input is shown in Fig.~\ref{Fsample_AVS}. Specifically, we assign the action features extracted by OpenPose and optical flow to the R channel, the slide transition features to the G channel, and the spectrogram voice features to the B channel. This method enables us to use well-established deep learning models like VGG-16, VGG-19, ResNet-50 and ResNet-101, which have proven effective in image classification tasks. The primary advantage of this approach is its computational efficiency - by processing all features through a single network path, we can significantly reduce memory usage and training time compared to parallel processing approaches. However, this integration introduces a notable limitation: the compression of feature maps into single-channel images leads to information loss. This is particularly problematic for action features, where lines in different colors represent distinct movement trajectories of tracked corner points. When these colored trajectories are compressed into a single channel, the spatial and temporal relationships between different movement patterns may become less distinguishable, potentially degrading the model's ability to learn complex motion patterns. The full architecture of the feature-level fusion strategy is illustrated in Fig.~\ref{feature_stacking}.

\begin{figure}[htbp]
\centering
\fbox{
\includegraphics[width=14 cm]{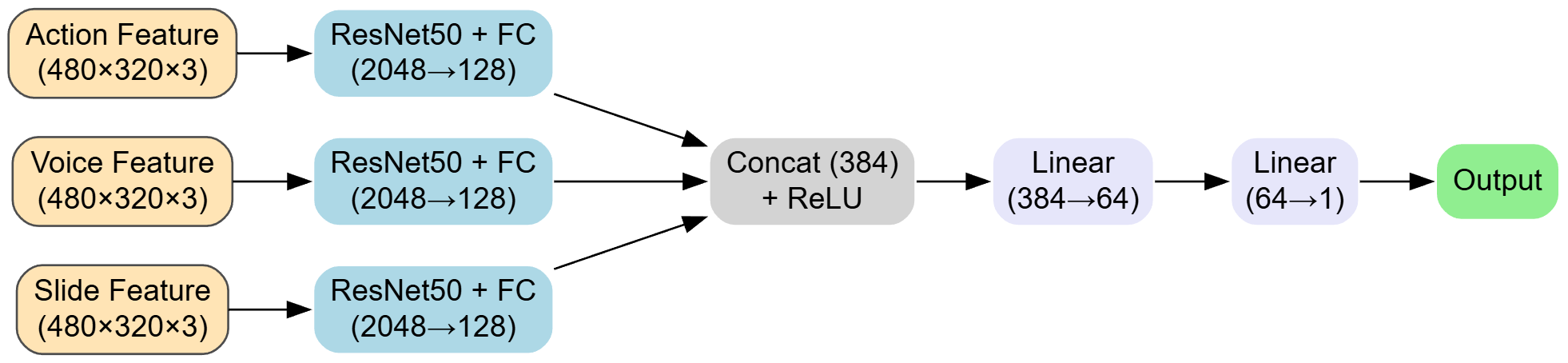}
}
\caption{Process of model-level fusion (example based on ResNet).}
\label{model_stacking}
\end{figure}

Another option, called "Model Stacking," employs multiple parallel deep learning models to process different input features independently before combining, as shown in Fig.~\ref{model_stacking}. In this architecture, each feature type (action, slide, and voice) is processed by its own dedicated neural network, preserving the complete dimensionality and characteristics of each feature type. The outputs from these individual networks are then concatenated at their fully connected layers to produce a final prediction. While this strategy maintains the complete information of each feature map and potentially allows for feature-specific network optimization, it comes with increased computational costs. These trade-offs will be further evaluated in the experiment section, where we compare the performance and efficiency of this approach with alternative architectures.

\subsubsection{Backbone Model Selection}

After confirming the superiority of feature-level fusion over model-level fusion in earlier experiments, we conduct all subsequent model evaluations under the feature-level fusion setting. Specifically, we compare several representative backbone architectures to identify the most suitable model for our regression task.

The evaluated backbones include:

\begin{itemize}
    \item \textbf{VGG-based models:} VGG16 and VGG19, known for their simplicity and deep convolutional stacks without residual connections.\\
    \item \textbf{Residual networks:} ResNet50 and ResNet101, which introduce skip connections to enable deeper and more stable training.\\
    \item \textbf{Transformer-based model:} Vision Transformer (ViT-16), which leverages self-attention to capture global dependencies.\\
    \item \textbf{Temporal model:} CNN+LSTM, combining spatial feature extraction and sequential modeling.
\end{itemize}

All models take as input a fused feature map constructed by stacking action, voice and slide modality features into a 3-channel image (480×320×3), and share a common training configuration. The evaluation is carried out using both regression metrics (MSE, MAE, $R^2$, PCC) and the accuracy of the 3-classification.

\subsubsection{Experimental Protocol}

All models were trained and evaluated using consistent procedures to ensure fair comparison across architectures and fusion strategies. The dataset consisted of seven lecture sessions. For the initial baseline experiment involving only action features and temporal smoothing, we used a fixed split: Lesson 1–5 for training, Lesson 6 for validation, and Lesson 7 for testing. All subsequent experiments—including multimodal fusion, ablation study and backbone comparisons—adopted a 7-fold cross-validation protocol at the lesson level. In each fold, one lesson was used for testing, while the remaining six were used for training and validation.

We employed the Adam optimizer with a fixed learning rate of $1 \times 10^{-5}$ and used the mean squared error (MSE) as the loss function. The batch size was set to 16, and the maximum number of training epochs was 200. Early stopping was applied with a patience of 25 epochs based on validation loss. A fixed random seed of 42 was used to ensure reproducibility.

All experiments were implemented in PyTorch and conducted on a workstation running Ubuntu 24.04. The system was equipped with an Intel Core i9-12900K processor, 128GB DDR4 RAM, and an NVIDIA RTX A6000 GPU.

\section{Experiment}
This section is currently under development and will be completed in a subsequent version.

\section{Discussion}
To be completed.

\section{Conclusion}
To be completed.


\begin{thebibliography}{99}

\bibitem{benedetto2024abstractive}
Benedetto, I., La Quatra, M., Cagliero, L., Canale, L., and Farinetti, L.  
\newblock Abstractive video lecture summarization: applications and future prospects.  
\newblock \textit{Education and Information Technologies}, 29(3):2951--2971, 2024.

\bibitem{sablic2021video}
Sabli{\'c}, M., Mirosavljevi{\'c}, A., and {\v{S}}kugor, A.  
\newblock Video-based learning (VBL)—past, present and future.  
\newblock \textit{Technology, Knowledge and Learning}, 26(4):1061--1077, 2021.

\bibitem{alaa2024video}
Alaa, T., Mongy, A., Bakr, A., Diab, M., and Gomaa, W.  
\newblock Video Summarization Techniques: A Comprehensive Review.  
\newblock \textit{arXiv preprint arXiv:2410.04449}, 2024.

\bibitem{han2019review}
Han, Z., Zhao, J., Leung, H., Ma, K.F., and Wang, W.  
\newblock A review of deep learning models for time series prediction.  
\newblock \textit{IEEE Sensors Journal}, 21(6):7833--7848, 2019.

\bibitem{singh2024bayesian}
Singh, A., and Kumar, M.  
\newblock Bayesian fuzzy clustering and deep CNN-based automatic video summarization.  
\newblock \textit{Multimedia Tools and Applications}, 83(1):963--1000, 2024.

\bibitem{10.1145/3661725.3661781}
Tan, K., Zhou, Y., Xia, Q., Liu, R., and Chen, Y.  
\newblock Large Model based Sequential Keyframe Extraction for Video Summarization.  
\newblock In \textit{Proceedings of the International Conference on Computing, Machine Learning and Data Science (CMLDS '24)}, pp. 52:1–5, 2024.

\bibitem{khan2024deep}
Khan, H., Hussain, T., Khan, S.U., Khan, Z.A., and Baik, S.W.  
\newblock Deep multi-scale pyramidal features network for supervised video summarization.  
\newblock \textit{Expert Systems with Applications}, 237:121288, 2024.

\bibitem{saykili2018distance}
Saykili, A.  
\newblock Distance education: Definitions, generations and key concepts and future directions.  
\newblock \textit{International Journal of Contemporary Educational Research}, 5(1):2--17, 2018.

\bibitem{vermeulen2024promoting}
Vermeulen, E.J., and Volman, M.L.L.  
\newblock Promoting student engagement in online education: Online learning experiences of Dutch university students.  
\newblock \textit{Technology, Knowledge and Learning}, 29(2):941--961, 2024.

\bibitem{deng2024effects}
Deng, R., and Gao, Y.  
\newblock Effects of embedded questions in pre-class videos on learner perceptions, video engagement, and learning performance.  
\newblock \textit{Active Learning in Higher Education}, 25(3):473--487, 2024.

\bibitem{kim2014understanding}
Kim, J., Guo, P.J., Seaton, D.T., Mitros, P., Gajos, K.Z., and Miller, R.C.  
\newblock Understanding in-video dropouts and interaction peaks in online lecture videos.  
\newblock In \textit{Proceedings of the First ACM Conference on Learning@Scale}, pp. 31--40, 2014.

\bibitem{guo2014video}
Guo, P.J., Kim, J., and Rubin, R.  
\newblock How video production affects student engagement: An empirical study of MOOC videos.  
\newblock In \textit{Proceedings of the First ACM Conference on Learning@Scale}, pp. 41--50, 2014.

\bibitem{bulathwela2020vlengagement}
Bulathwela, S., Perez-Ortiz, M., Yilmaz, E., and Shawe-Taylor, J.  
\newblock VLEngagement: A dataset of scientific video lectures for evaluating population-based engagement.  
\newblock \textit{arXiv preprint arXiv:2011.02273}, 2020.

\bibitem{dimitrova2004context}
Dimitrova, N.  
\newblock Context and memory in multimedia content analysis.  
\newblock \textit{IEEE Multimedia}, 11(3):7--11, 2004.

\bibitem{agyeman2019soccer}
Agyeman, R., Muhammad, R., and Choi, G.S.  
\newblock Soccer video summarization using deep learning.  
\newblock In \textit{Proc. 2019 IEEE MIPR}, pp. 270--273, 2019.

\bibitem{andra2019automatic}
Andra, M.B., and Usagawa, T.  
\newblock Automatic lecture video content summarization with attention-based recurrent neural network.  
\newblock In \textit{Proc. 2019 ICAIIT}, pp. 54--59, 2019.

\bibitem{chen2015effects}
Chen, C.M., and Wu, C.H.  
\newblock Effects of different video lecture types on attention, emotion, cognitive load, and learning performance.  
\newblock \textit{Computers \& Education}, 80:108--121, 2015.

\bibitem{zhang2018effects}
Zhang, J., Bourguet, M.L., and Venture, G.  
\newblock Effects of video instructor's body language on students' visual attention: An eye-tracking study.  
\newblock In \textit{Proc. 32nd International BCS HCI Conference}, pp. 1--5, 2018.

\bibitem{burton1978thinking}
Burton, A., and Radford, J.  
\newblock \textit{Thinking in Perspective: Critical Essays in the Study of Thought Processes}.  
\newblock Routledge, 1978.

\bibitem{cao2019openpose}
Cao, Z., Hidalgo, G., Simon, T., Wei, S.E., and Sheikh, Y.  
\newblock OpenPose: Real-time multi-person 2D pose estimation using Part Affinity Fields.  
\newblock \textit{IEEE Transactions on PAMI}, 43(1):172--186, 2019.

\bibitem{lucas1981iterative}
Lucas, B.D., and Kanade, T.  
\newblock An iterative image registration technique with an application to stereo vision.  
\newblock In \textit{Proc. IJCAI}, 1981.

\bibitem{wyse2017audio}
Wyse, L.  
\newblock Audio spectrogram representations for processing with convolutional neural networks.  
\newblock \textit{arXiv preprint arXiv:1706.09559}, 2017.

\bibitem{flanagan1972speech}
Flanagan, J.L.  
\newblock Speech synthesis.  
\newblock In \textit{Speech Analysis Synthesis and Perception}, pp. 204--276, Springer, 1972.

\bibitem{slykhuis2005eye}
Slykhuis, D.A., Wiebe, E.N., and Annetta, L.A.  
\newblock Eye-tracking students' attention to PowerPoint photographs in science education.  
\newblock \textit{Journal of Science Education and Technology}, 14(5):509--520, 2005.

\bibitem{sola1997importance}
Sola, J., and Sevilla, J.  
\newblock Importance of input data normalization for neural networks in industrial problems.  
\newblock \textit{IEEE Trans. on Nuclear Science}, 44(3):1464--1468, 1997.

\bibitem{press1990savitzky}
Savitzky, A., and Golay, M.J.E.  
\newblock Smoothing and differentiation of data by simplified least squares procedures.  
\newblock \textit{Analytical Chemistry}, 36(8):1627--1639, 1964.

\bibitem{abuhammad2020barriers}
Abuhammad, S.  
\newblock Barriers to distance learning during the COVID-19 outbreak: A qualitative review.  
\newblock \textit{Heliyon}, 6(11), 2020.

\bibitem{hasegawa2007}
Hasegawa, S., Tajima, Y., Matou, M., Futatsudera, M., and Ando, T.  
\newblock Case studies for self-directed learning using lecture archives.  
\newblock In \textit{Proc. WBE 2007}, pp. 299--304, 2007.

\bibitem{vimalaksha2018automated}
Vimalaksha, A., Vinay, S., Prekash, A., and Kumar, N.S.  
\newblock Automated summarization of lecture videos.  
\newblock In \textit{Proc. IEEE T4E}, pp. 126--129, 2018.

\bibitem{shi2024impacts}
Shi, Y., Wang, M., Chen, Z., Hou, G., Wang, Z., Zheng, Q., and Sun, J.  
\newblock The impacts of instructor’s visual attention and lecture type on learning performance.  
\newblock \textit{Education and Information Technologies}, 2024.

\bibitem{simonyan2014very}
Simonyan, K.  
\newblock Very deep convolutional networks for large-scale image recognition.  
\newblock \textit{arXiv preprint arXiv:1409.1556}, 2014.

\end{thebibliography}
\end{document}